\documentclass[10pt,letterpaper]{article}
\usepackage{opex3}
\usepackage{amsmath,amssymb,graphicx}
\usepackage{epstopdf}

\begin{document}

\title{Active optomechanics through relaxation oscillations}

\author{Debora Princepe, Gustavo Wiederhecker and Newton Frateschi}

\address{“Gleb Wataghin” Physics Institute, University of Campinas, Campinas-SP, 13083-859, Brazil}
\email{princepe@ifi.unicamp.br}

\begin{abstract}We propose an optomechanical laser based on III-V compounds which exhibits self-pulsation in the presence of a dissipative optomechanical coupling. In such a laser cavity, radiation pressure drives the mechanical degree of freedom and its back-action is caused by the mechanical modulation of the cavity loss rate. Our numerical analysis shows that even in a wideband gain material, such dissipative coupling couples the mechanical oscillation with the laser relaxation oscillations process. Laser self-pulsation is observed for mechanical frequencies below the laser relaxation oscillation frequency under sufficiently high optomechanical coupling factor. 
\end{abstract}

\section{Introduction}
The coupling between light and mechanical vibrations in optical microcavities usually occurs through a  dispersive coupling between the optical resonant frequency and the mechanical vibration. Such dispersive scheme, where the cavity frequency shifts due to the mechanical motion, has been used to demonstrate amplification, cooling and interference of mechanical modes \cite{aspelmeyer_cavity_2013, kippenberg_cavity_2008, chan_laser_2011,zhang_synchronization_2012, lin_mechanical_2009,dobrindt_parametric_2008}. More recently several theoretical studies have explored optomechanical interaction within an optically active medium, which may relax the so-called resolved side-band regime, necessary to achieve ground state cooling~\cite{ge_gain-tunable_2013,genes_micromechanical_2009,usami_optical_2012}. In contrast with passive optomechanical cavities, active optomechanical cavities exhibiting purely dispersive optomechanical exhibit optomechanical back-action either under an external coherent injection signal laser \cite{ge_gain-tunable_2013} or when the gain material has a narrowband gain \cite{genes_micromechanical_2009}. Active cavities possessing mechanical degrees of freedom have been demonstrated experimentally in the context of tunable micro lasers \cite{huang_nanoelectromechanical_2008,yang_linewidth_2013,czerniuk_lasing_2014}. It has been shown that the cavity length fluctuations in a vertical cavity surface emitting laser (VCSEL) lead to a broader emission linewidth \cite{yang_linewidth_2013}. The impact of strain waves over the light emitted by a laser cavity has also been explored and reveals the complex interaction between phonons, photons and charge carriers in active materials \cite{czerniuk_lasing_2014,fainstein_strong_2013,yeo_strain-mediated_2014,xuereb_exciton-mediated_2012}.

Here we propose an optomechanical device based on the broadband gain III-V materials and a dissipative optomechanical coupling –- essentially the mechanical displacement leads to a modulation of the cavity optical loss rate. It is shown that such combination leads to a novel coupling between the mechanical mode and the relaxation oscillations of the laser cavity. 

\section{Mechanically induced loss modulation}

\begin{figure}[tbp]
\centerline{\includegraphics[scale=0.6]{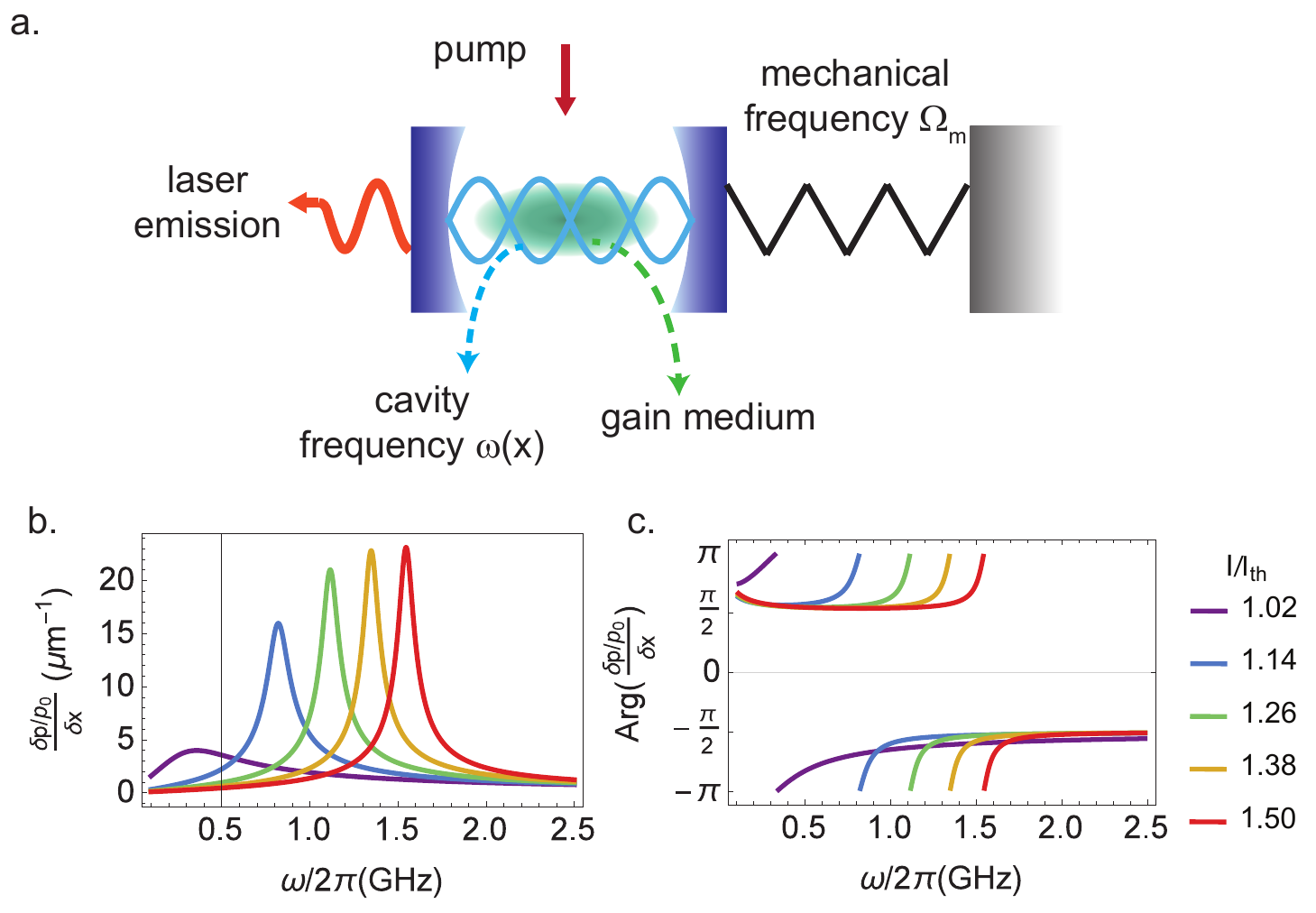}}
\caption{\label{fig:schematic}(a) Scheme of the proposed system: a laser cavity with a mechanical degree of freedom for modulation of the emitted light. (b) Relative photon density fluctuation $\delta p (\omega)/(p_0 \delta x)$ with variable frequency $\omega$ of the signal response, for different values of current, multiples of the laser threshold current $I_{th}$. The maximum response always occurs for frequency equal to the relaxation oscillation frequency (ROF, $\Omega_r$). The inset shows the relative phase of the response, where for $\omega<\Omega_r$ ($\omega>\Omega_r$) fluctuations in $p$ lead (lag) fluctuations in $x$.}
\label{fig:fig1}
\end{figure}

The laser-cavity response to the mechanical motion is investigated through a displacement-dependent optical resonance, as illustrated in the schematic shown in Fig. \ref{fig:schematic}. The optical frequency is written as $\omega(x)=\omega_0+g_{\omega}x(t)$, where $\omega_0$ is the bare cavity optical frequency and $g_{\omega}=d\omega/dx$ is the usual dispersive optomechanical coupling. Such dispersive coupling naturally leads to a position dependent cavity loss rate, $\kappa(x)=\omega(x)/Q_0$ where $Q_0$ is the intrinsic optical quality factor. For example, in a Fabry-Perot cavity with length $L(x)$, the cavity loss rate is given by $\kappa(x)= v_g/(L(x)\ln(1/R))$ where $v_g$ is the group velocity of light in the material and $R$ is the mirror reflectivity. Therefore it is convenient to define the dissipative optomechanical coupling rate as $\kappa(x)=\kappa_0+g_{\kappa}x(t)$ with $g_{\kappa}=g_{\omega}/Q_0$. In other words, once we admit a mechanical degree of freedom to an optical cavity, its loss is modulated by the mechanical oscillation, although the optical force has a dispersive nature, generated by fluctuation of the total cavity energy. In passive high-Q cavities the dissipative coupling is rather small, however active cavities display a strong intrinsic absorption and can exhibit a reasonable dissipative coupling rate. Indeed it has been recently demonstrated that the mechanical Brownian motion leads to linewidth broadening in VCSEL lasers \cite{yang_linewidth_2013}.

The laser cavity response to a small amplitude oscillation of the mechanical degree of freedom can be derived from the well-known laser rate equations  \cite{agrawal_semiconductor_1993} for the photon number density $p$ and the excess carrier concentration $n$, 
\begin{eqnarray}
\label{eq:eq1}
\frac{dn}{dt}=\frac{I}{qV}-R_{sp}-Gp\\
\label{eq:eq2}
\frac{dp}{dt}=[G-\gamma]p-\kappa(x) p+\beta R_{sp}
\end{eqnarray}
where $I$ is the injected carrier flux obtained either electrically or optically, $q$ is the elementary charge, $V$ is the cavity volume, $R_{sp} = Bn^2$  is the spontaneous emission rate; $G$ is the carrier-density dependent stimulated emission rate; $\gamma$ is the intrinsic loss rate; $\beta$ is the fraction of the spontaneous emission coupled to the optical field. To calculate the impact of mechanically induced fluctuations on the laser dynamics we assume a small amplitude oscillation for $x(t)$, $x(t)=x_0+\delta x e^{-i \omega t}$, where $x_0$ is the static displacement, and sought for the carrier and photon density response using the ansatz, $n(t)=n_0+\delta n e^{-i \omega t}$ and $p(t)=p_0+\delta p e^{-i \omega t}$, where $n_0$ and $p_0$ are the steady state solutions which represent the average carrier concentration and intracavity photon density. The relative photon density fluctuation $\delta p/(p_0 \delta x)$ represents the optical modulation depth displacement sensitivity. Within the small signal approximation the displacement induced modulation is given by,
\begin{eqnarray}
\label{eq:eq3}
\frac{\delta p}{p_0 \delta x}(\omega)=-g_{\kappa}\frac{\Gamma_n - i \omega}{\Omega_r^2-(\omega+i\Gamma_r)^2}
\end{eqnarray}
where $\Gamma_n=2Bn_0+G_n p_0$ is carrier damping rate, $\Omega_r^2=G_0G_np_0 + 2G_0\beta Bn_0-(\Gamma_n -\Gamma_p)^2/4$, is the usual relaxation oscillation frequency (ROF), with a decay rate $\Gamma_r=(\Gamma_n+\Gamma_p)/2$ and the photon fluctuation decay rate is given by $\Gamma_p=\beta R_{sp}/p_0$. The carrier-density stimulated rate is linearized 
as $G(n)=G_n (n-n_{tr})$, where $G_n = dG/dn$ is the differential gain and $n_{tr}$ is the carrier density at transparency; $G_0$ is the gain value at steady-state \cite{agrawal_semiconductor_1993}. The modulation depth frequency response is shown Fig. \ref{fig:schematic}(b) for several injection currents in a typical III-V microdisk laser with the following parameters: $(V, B, G_n, n_{tr}, \gamma, Q_0, \beta) = (1.3 \times 10^{-16},10^{-16},1.3 \times 10^{-12}, 7 \times 10^{23},4.3 \times 10^{11},5000,10^{-4})$ at SI units \cite{agrawal_semiconductor_1993}. The typical carrier and photon rates  are obtained from the steady state solution slightly above the laser threshold. For a current $20\%$ higher than the threshold, $I=1.2\; I_{th}$,  we obtain $(\Gamma_n,\Gamma_p,\Omega_r,\Gamma_r)/2\pi=(48, 52, 980, 50)$MHz -- this current was fixed in our calculations. For higher currents the value of $\Omega_r$ increases and $\Gamma_r$ decreases. The optomechanical dispersive coupling was estimated as $g_{\omega}/2\pi\approx -10$ GHz/nm, based on typical semiconductor microdisks in the literature \cite{zhang_synchronization_2012,mitchell_cavity_2014}, therefore  $g_{\kappa}\approx-2 (\times 2\pi)$ MHz/nm. The strong resonance enhancement of the modulation depth shown in Fig. \ref{fig:schematic}(b) occurs when the mechanical frequency matches the laser ROF, $\omega=\Omega_r$, implying that the photon density fluctuations may be strongly amplified. Interestingly, the phase response implies that  $\delta p$ leads (lags) $\delta x$ when $\omega<\Omega_r$ ($\omega>\Omega_r$), i.e., the mechanical frequency is red-detuned (blue-detuned) relative to the ROF. 

\section{Optical spring effect and self-sustained oscillations}
\begin{figure}[tbp]
\centerline{\includegraphics[scale=0.45]{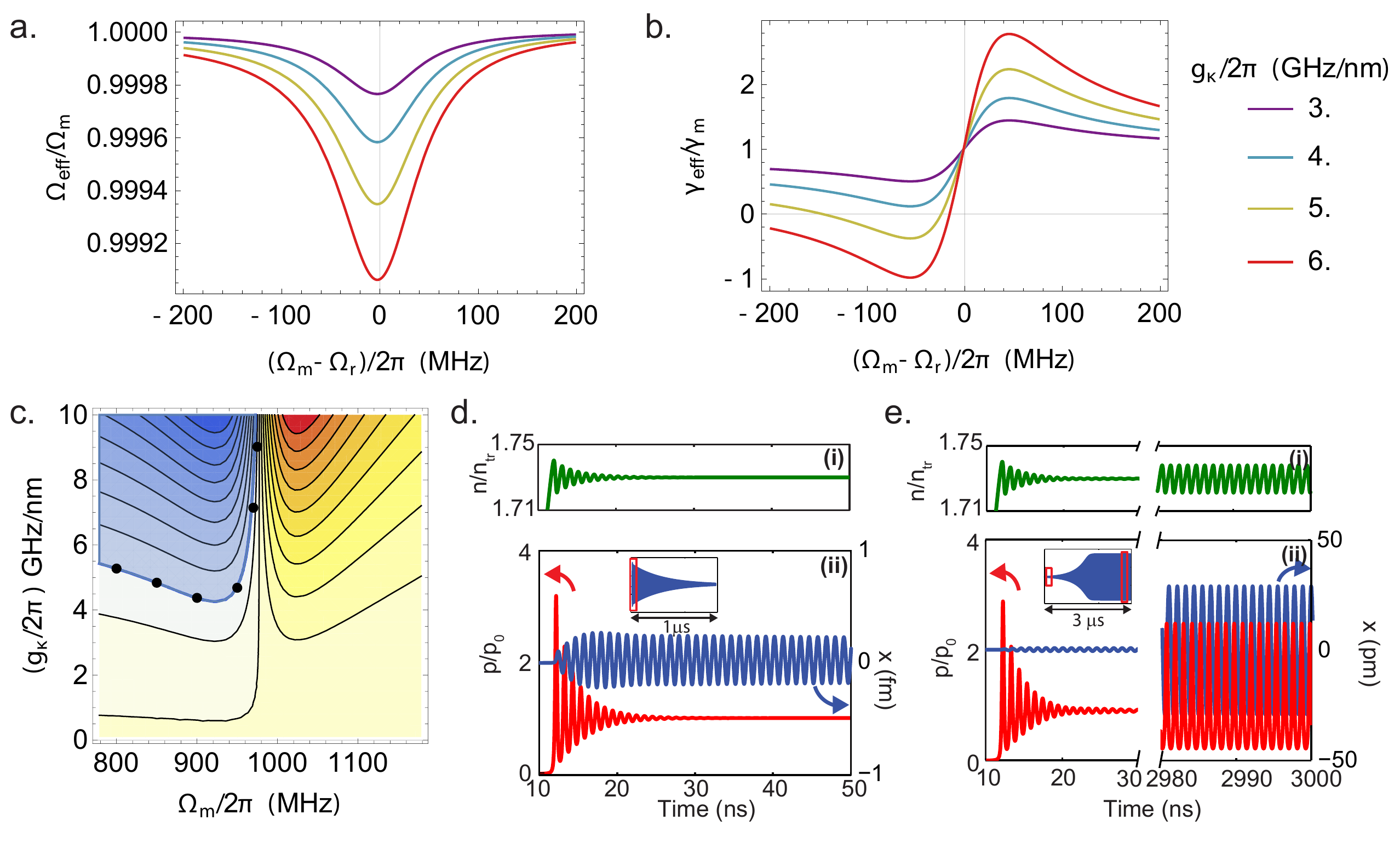}}
\caption{(a) Relative optical spring effect and (b) optomechanical damping for $I=1.2\;I_{th}$. (c) Contour map of $\gamma_{eff}$ for different values of $\Omega_m$ and coupling factor $g_{\kappa}$. The blue region corresponds to $\gamma_{eff}<0$, i.e., the amplification region. The dots correspond to the dynamic simulation, where the behaviour changes from damped to amplified oscillations, shown in (d) and (e) respectively. In (d) $g_{\kappa}=-2$MHz/nm and in (e) $g_{\kappa}=-6$GHz/nm  The mechanical behaviour is reproduced in carrier (i - green line) and photon fluctuations (ii - red line), due to the coupling between relaxation oscillations and mechanical motion (ii - blue line). The insets show the mechanical evolution after a long period of time.}
\label{fig:spring_damping}
\end{figure}

The resonant enhancement of the optical modulation depth surrounding the relaxation oscillation frequency indicates that an oscillating optical force will drive the mechanical degree of freedom. We investigated the effects of this optical force by considering a driven harmonic mechanical oscillator,
\begin{eqnarray}
\label{eq:eq4}
\frac{d^2x}{dt^2}+\gamma_m \frac{dx}{dt}+\Omega_m^2 x=-\frac{\hbar P Q_0 g_{\kappa}}{m_{eff}}
\end{eqnarray}
where $\Omega_m$ is the mechanical resonance frequency, the mechanical damping rate is $\gamma_m=\Omega_m/Q_m$, $m_{eff}$ is the effective motional mass and $P=pV$ is the  total intracavity photon number. Using the ansatz $x(t)=x_0+\delta x e^{-i \omega t}$ in Eq. (\ref{eq:eq4}) together with Eq. (\ref{eq:eq3}) for the photon modulation, we can arrive in a harmonic oscillator equation for the mechanical fluctuation with modified frequency and damping terms,
\begin{eqnarray}
\label{eq:eq5}
\delta \Omega_m(\omega) = -g^2 Q_0 \frac{\Omega_m}{\omega}\frac{\Gamma_n (\Omega_r^2 -\omega ^2+\Gamma_r^2)+2 \Gamma_r \omega^2}{(\Gamma_r^2+\omega^2)^2+2 \Omega_r^2 (\Gamma_r-\omega)(\Gamma_r+\omega)+\Omega_r^4},\\
\label{eq:eq6}
\gamma_{opt}(\omega)= -g^2 Q_0 \frac{\Omega_m}{\omega} \frac{2\omega( \Omega_r^2-\omega^2+\Gamma_r^2 - 2\Gamma_n \Gamma_r)}{(\Gamma_r^2+\omega^2)^2 + 2 \Omega_r^2 (\Gamma_r-\omega)(\Gamma_r+\omega)+\Omega_r^4},
\end{eqnarray}
where the effective dissipative coupling was defined as $g=g_{\kappa}x_\text{zpf}\sqrt{P_0}$ with $x_\text{zpf}=\sqrt{\hbar/(2 m_{eff}\Omega_m)}$ \cite{aspelmeyer_cavity_2013}. In analogy with passive optomechanical cavities the origin of the optical spring effect (Eq. (\ref{eq:eq5})) and optical damping (Eq.(\ref{eq:eq6})) lies on the phase lag between the optical forces induced by the fluctuation optical intensity and the mechanical motion. Here however the optical intensity fluctuations are caused by a dissipative process and has an effect similar to a saturable absorber inside a laser cavity, where the intensity driven displacement modulates the cavity loss. In Fig. \ref{fig:spring_damping} we illustrate the predicted optical spring (Fig. \ref{fig:spring_damping}(a)) and damping (Fig. \ref{fig:spring_damping}(b)) calculated using Eq.(\ref{eq:eq5}) and (\ref{eq:eq6}), omitting the minus sign of $g_{\kappa}$. In contrast to the passive optomechanical cavities, where the spring effect can be both positive and negative, the relaxation oscillation induced spring can only soften the mechanical spring. This is expected for a purely dissipative coupling since the mechanical motion modulates the cavity loss regardlessly of the relative detuning of between the ROF and the mechanical frequency.  Differently, the optical damping, which is related to the lag in the laser response, can be positive or negative depending on the ROF detuning with respect to the mechanical frequency.
Hereafter we focused on the red-detuned ($\Omega_m<\Omega_r$) regime that may amplify the mechanical vibrations. Fig. \ref{fig:spring_damping}(b) indicates that at large $g_\kappa$ the system may exhibit a negative damping and initiate self-sustained oscillations.  From Eq. (\ref{eq:eq6}), we can predict a threshold for the build-up of these oscillations, when the effective damping $\gamma_{eff}=\gamma_{opt} + \gamma_m$ becomes negative, i. e. the condition $\gamma_{eff}=0$ is satisfied.
Using Eq. (\ref{eq:eq6}) and the same typical laser parameters used in Fig. \ref{fig:schematic}(b), and an harmonic oscillator with $m_{eff}$=50 pg and $Q_m$=1000, we conclude that $g_\kappa\approx -6$ GHz/nm is sufficient to achieve self-sustained oscillations for such value of pump current.

We verified that the small-signal analysis is consistent by performing a full numerical integration of the laser rate equations (Eqs. \ref{eq:eq1},\ref{eq:eq2}) coupled to the mechanical harmonic oscillator (Eq. \ref{eq:eq4}). In Fig. \ref{fig:spring_damping}(c) a contour plot of the effective mechanical damping, as given by Eq. \ref{eq:eq6},  is shown as a function of the dissipative coupling rate $g_\kappa$. The $g_\kappa-\Omega_m$ region corresponding to negative damping ($\gamma_\text{eff}<0$) is highlighted in blue. The black points correspond to the onset of self-sustained oscillations obtained from the full numerical simulations and they are in very good agreement with the analytically calculated thresholds.  The dynamical behaviour for two distinct values of $g_k$ is shown in Fig.\ref{fig:spring_damping}(d,e) for a $\Omega_m/(2 \pi)=900$ MHz mechanical oscillator.
In either case the relaxation oscillations of the laser decays with time and the carrier dynamics apparently reaches a steady state. For a $g_\kappa/(2 \pi)=-2$MHz/nm, outside the instability region in Fig \ref{fig:spring_damping}(d), the mechanical oscillator receives an initial kick after the laser is turned on, around $t\approx10 ns$, but decays to a stationary state on a time scale $\tau_m\approx Q_m/\Omega_m$. In Fig. \ref{fig:spring_damping}(e) however, after the initial kick the mechanical oscillation amplitude grows on a time scale $\tau_m$ until it reaches a dynamical steady-state. At this dynamical steady-state one can notice that both the carrier and photon population are modulated at mechanical oscillator frequency.  Although the $g_{\kappa}$ needed to achieve this simulated regime is still very large compared to what typical optomechanical systems usually provide -- $g_\kappa/(2 \pi)=-6$GHz/nm, this novel interaction between the carriers, photons and the mechanical oscillator lead us to investigate different materials and new geometries for the experimental observation of the phenomena -– III-V-based single nanodisk have already been demonstrated and achieved experimental higher values of $g_{\kappa}$ than usual systems demonstrated \cite{ding_high_2010,baker_photoelastic_2014}.

\section{Normal mode splitting}

\begin{figure}[b]
\centerline{\includegraphics[scale=0.5]{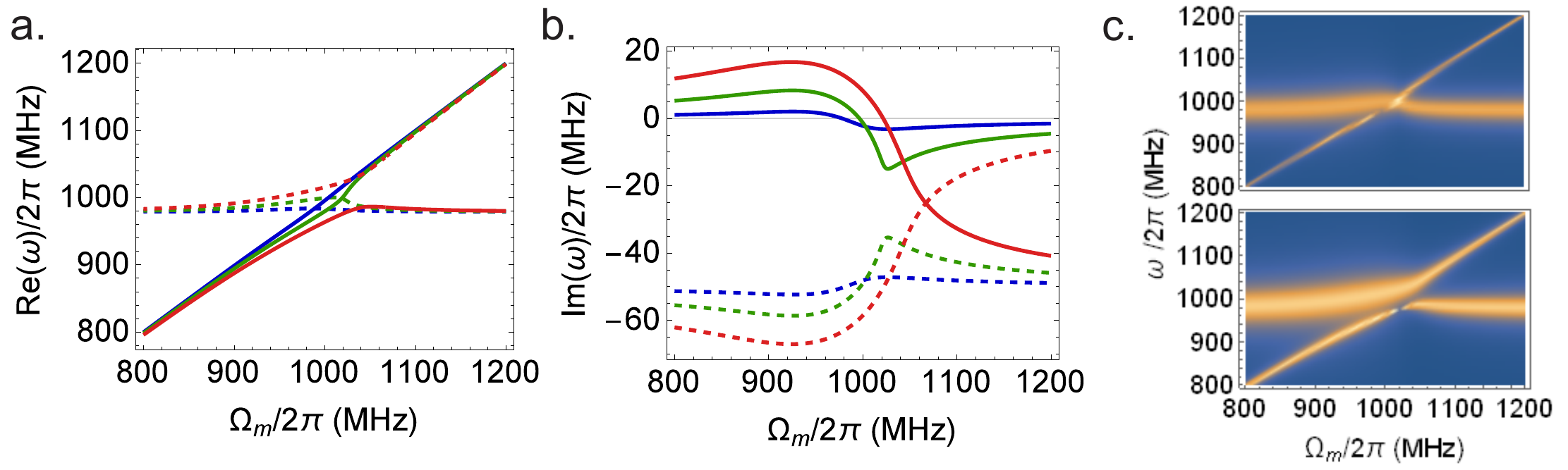}}
\caption{(a)Real and (b) imaginary parts of the solutions of $\omega$ for $g_{\kappa} = 10$, $20$ and $30$GHz/nm -- blue, green and red lines respectively. Full lines correspond to the effective mechanical response and dashed lines correspond to effective relaxation oscillation. (c)The density plot shows the response frequencies with thickness given by each effective damping, for $g_\kappa$ equal to $20$ (top) and $30$GHz/nm (bottom). The blank point corresponds to zero linewidth, i.e., $\gamma_{eff}=0$.}
\label{fig:eigenvalues}
\end{figure}

The linearized results obtained previously are valid for weak dissipative coupling. In order to investigate beyond this regime, we calculated the  response frequency $\omega$ from solving the coupled equations (\ref{eq:eq1}, \ref{eq:eq2}, \ref{eq:eq4}):
\begin{eqnarray}
\label{eq:eq7}
\frac{\Omega_r ^2 - (\omega+i\Gamma_r)^2}{\Gamma_n - i \omega}-\frac{2g^2Q_0\Omega_m}{\Omega_m^2-\omega^2-i\gamma_m \omega}=0
\end{eqnarray}
This equation has two pairs of conjugated solutions. The real part of each root is related either to the mechanical frequency or to the ROF; the imaginary part determines the stability of the solutions and they are related to $\gamma_m/2$ and $\Gamma_r$. The real and the imaginary parts of the two complex roots are shown in Fig. \ref{fig:eigenvalues}(a,b) as function of the mechanical frequency for several values of $g_{\kappa}$. In Fig. \ref{fig:eigenvalues}(a) the interaction between the mechanical oscillator (solid curves) and the ROF (dashed curves) is noticeable around their crossing point ($\Omega_m/2\pi\approx 980$ MHz). In agreement with the small-signal analysis, the dispersion of the eigenvalue related to the mechanical oscillation is red-shifted, while the relaxation oscillation frequency is blue-shifted,  as a result of the crossing between these oscillators. 
The imaginary parts of the eigenvalues show that the damping of the mechanical oscillator is affected by the optomechanical damping such that it is possible the imaginary part becomes positive, that indicates the amplification of the mechanical oscillation. The choose of parameters was such that $\gamma_m$ was much lower than $\Gamma_r$ (at 980MHz, $\gamma_m=0.98$MHz and $\Gamma_r=50$MHz), then the last one is less affected. The symmetry of the solutions is compatible with the problem of coupling two damped oscillators with the presence of gain. This problem has been explored recently showing interesting effects such as changes of the imaginary part of the frequency response and non intuitive behaviour after laser threshold\cite{jing_pt-symmetric_2014,brandstetter_reversing_2014}. In Fig. \ref{fig:eigenvalues}(c) we present the real part of the solutions with thickness proportional to each imaginary part. For the condition of high coupling, such as $(g_{\omega}x_{zpf}) \gtrsim\Gamma_r, \gamma_m$ , there's an anti-crossing between the solutions, typical of strong coupling \cite{dobrindt_parametric_2008,groblacher_observation_2009}. We observe that at this condition also $\Gamma_r$ is modified. The point where the mechanical solution is blanked corresponds to the condition where $\gamma_{eff}=0$. For week coupling, this only occurs for $\Omega_m<\Omega_r$, but for strong coupling it occurs around the region of anti-crossing, due to the bigger change in the relaxation oscillation frequency. In either case, we identify, as predicted, the resonance of $\Omega_r$ and $\Omega_m$, where the first act as a cut off frequency for the amplification effect.

This treatment is not adequate for the investigation of the ``blue side'' of the optomechanical laser ($\Omega_m>\Omega_r$), where the imaginary part is negative; since we employed a semi classical approach, we haven't included thermal bath which is necessary for the verification of the cooling of the mechanical oscillation in this region. Nevertheless, previous theoretical work with appropriate quantum mechanics treatment shows that a laser cavity is inefficient in the process of cooling its own mirror when the pump is incoherent, and the process only occurs efficiently in the presence of coherent pump \cite{ge_gain-tunable_2013}. Our approach is suitable for the condition of incoherent pump, since injected carriers generate spontaneous emission, therefore we don't expect this system to have efficient cooling. 

\section{Conclusion}
We have shown the existence of coupling between the laser relaxation oscillation, slightly above the threshold, and the mechanical mode, in a dissipative optomechanical resonator based on a material with optical gain. Small signal analysis shows that there's a minimum coupling to achieve self-sustained oscillation near the steady state that could lead to modulation of the emitted light. The effect only occurs when $\Omega_m<\Omega_r$. Although the optomechanical coupling factor required to achieve this condition is higher than simulated for our devices, we expect the model to be valid for other geometries. This leads us to investigate different systems in which the minimum condition is more easily achieved.

\section*{Acknowledgments}
This work was supported by the Brazilian financial agencies: CNPq, S\~{a}o Paulo Research Foundation (FAPESP) under grant $\#2011/18945-6$ and National Institute for Science and Technology (FOTONICOM) under grant $\#08/57857-2$, S\~{a}o Paulo Research Foundation (FAPESP).


\begin{thebibliography}{10}
\newcommand{\enquote}[1]{``#1''}

\bibitem{aspelmeyer_cavity_2013}
M.~Aspelmeyer, T.~J. Kippenberg, and F.~Marquardt, \enquote{Cavity
  optomechanics,} {arXiv}:1303.0733  (2013).

\bibitem{kippenberg_cavity_2008}
T.~J. Kippenberg and K.~J. Vahala, \enquote{Cavity optomechanics: Back-action
  at the mesoscale,} Science \textbf{321}, 1172--1176 (2008).

\bibitem{chan_laser_2011}
J.~Chan, T.~P.~M. Alegre, A.~H. Safavi-Naeini, J.~T. Hill, A.~Krause,
  S.~Gr\"{o}blacher, M.~Aspelmeyer, and O.~Painter, \enquote{Laser cooling of a
  nanomechanical oscillator into its quantum ground state,} Nature
  \textbf{478}, 89--92 (2011).

\bibitem{zhang_synchronization_2012}
M.~Zhang, G.~S. Wiederhecker, S.~Manipatruni, A.~Barnard, P.~McEuen, and
  M.~Lipson, \enquote{Synchronization of micromechanical oscillators using
  light,} Physical Review Letters \textbf{109} (2012).

\bibitem{lin_mechanical_2009}
Q.~Lin, J.~Rosenberg, X.~Jiang, K.~J. Vahala, and O.~Painter,
  \enquote{Mechanical oscillation and cooling actuated by the optical gradient
  force,} Phys. Rev. Lett. \textbf{103}, 103601 (2009).

\bibitem{dobrindt_parametric_2008}
J.~M. Dobrindt, I.~Wilson-Rae, and T.~J. Kippenberg, \enquote{Parametric
  normal-mode splitting in cavity optomechanics,} Phys. Rev. Lett.
  \textbf{101}, 263602 (2008).

\bibitem{ge_gain-tunable_2013}
L.~Ge, S.~Faez, F.~Marquardt, and H.~E. T\"{u}reci, \enquote{Gain-tunable
  optomechanical cooling in a laser cavity,} Phys. Rev. A \textbf{87}, 053839
  (2013).

\bibitem{genes_micromechanical_2009}
C.~Genes, H.~Ritsch, and D.~Vitali, \enquote{Micromechanical oscillator
  ground-state cooling via resonant intracavity optical gain or absorption,}
  Phys. Rev. A \textbf{80}, 061803 (2009).

\bibitem{usami_optical_2012}
K.~Usami, A.~Naesby, T.~Bagci, B.~M. Nielsen, J.~Liu, S.~Stobbe, P.~Lodahl, and
  E.~S. Polzik, \enquote{Optical cavity cooling of mechanical modes of a
  semiconductor nanomembrane,} Nature Physics \textbf{8} (2012).

\bibitem{huang_nanoelectromechanical_2008}
M.~C.~Y. Huang, Y.~Zhou, and C.~J. Chang-Hasnain, \enquote{A
  nanoelectromechanical tunable laser,} Nat Photon \textbf{2}, 180--184 (2008).

\bibitem{yang_linewidth_2013}
W.~Yang, Y.~Rao, C.~Chase, M.~C. Huang, and C.~Chang-Hasnain,
  \enquote{Linewidth measurement of 1550 nm high contrast grating
  {MEMS}-{VCSELs},} in \enquote{{CLEO}: 2013,}  (Optical Society of America,
  2013), {OSA} Technical Digest (online), p. CF2F.4.

\bibitem{czerniuk_lasing_2014}
T.~Czerniuk, C.~Brüggemann, J.~Tepper, S.~Brodbeck, C.~Schneider, M.~Kamp,
  S.~H\"{o}fling, B.~A. Glavin, D.~R. Yakovlev, A.~V. Akimov, and M.~Bayer,
  \enquote{Lasing from active optomechanical resonators,} Nat Commun \textbf{5}
  (2014).

\bibitem{fainstein_strong_2013}
A.~Fainstein, N.~D. Lanzillotti-Kimura, B.~Jusserand, and B.~Perrin,
  \enquote{Strong optical-mechanical coupling in a vertical {GaAs}/{AlAs}
  microcavity for subterahertz phonons and near-infrared light,} Phys. Rev.
  Lett. \textbf{110}, 037403 (2013).

\bibitem{yeo_strain-mediated_2014}
I.~Yeo, P.-L.~d. Assis, A.~Gloppe, E.~Dupont-Ferrier, P.~Verlot, N.~S. Malik,
  E.~Dupuy, J.~Claudon, J.-M. Gérard, A.~Auffèves, G.~Nogues, S.~Seidelin,
  J.-P. Poizat, O.~Arcizet, and M.~Richard, \enquote{Strain-mediated coupling
  in a quantum dot-mechanical oscillator hybrid system,} Nat Nano \textbf{9},
  106--110 (2014).

\bibitem{xuereb_exciton-mediated_2012}
A.~Xuereb, K.~Usami, A.~Naesby, E.~S. Polzik, and K.~Hammerer,
  \enquote{Exciton-mediated photothermal cooling in {GaAs} membranes,} New J.
  Phys. \textbf{14}, 085024 (2012).

\bibitem{agrawal_semiconductor_1993}
G.~P. Agrawal and N.~K. Dutta, \emph{Semiconductor Lasers} (Kluwer Academic
  Pub, 1993).

\bibitem{mitchell_cavity_2014}
M.~Mitchell, A.~C. Hryciw, and P.~E. Barclay, \enquote{Cavity optomechanics in
  gallium phosphide microdisks,} Applied Physics Letters \textbf{104}, 141104
  (2014).

\bibitem{ding_high_2010}
L.~Ding, C.~Baker, P.~Senellart, A.~Lemaitre, S.~Ducci, G.~Leo, and I.~Favero,
  \enquote{High frequency {GaAs} nano-optomechanical disk resonator.} Physical
  review letters \textbf{105} (2010).

\bibitem{baker_photoelastic_2014}
C.~Baker, W.~Hease, D.-T. Nguyen, A.~Andronico, S.~Ducci, G.~Leo, and
  I.~Favero, \enquote{Photoelastic coupling in gallium arsenide optomechanical
  disk resonators,} Optics Express \textbf{22}, 14072 (2014).

\bibitem{jing_pt-symmetric_2014}
H.~Jing, S.~\"{O}zdemir, X.-Y. Lü, J.~Zhang, L.~Yang, and F.~Nori,
  \enquote{{PT}-symmetric phonon laser,} Phys. Rev. Lett. \textbf{113}, 053604
  (2014).

\bibitem{brandstetter_reversing_2014}
M.~Brandstetter, M.~Liertzer, C.~Deutsch, P.~Klang, J.~Sch\"{o}berl, H.~E.
  T\"{u}reci, G.~Strasser, K.~Unterrainer, and S.~Rotter, \enquote{Reversing the
  pump dependence of a laser at an exceptional point,} Nat Commun \textbf{5}
  (2014).

\bibitem{groblacher_observation_2009}
S.~Gr\"{o}blacher, K.~Hammerer, M.~R. Vanner, and M.~Aspelmeyer,
  \enquote{Observation of strong coupling between a micromechanical resonator
  and an optical cavity field,} Nature \textbf{460}, 724--727 (2009).

\end{thebibliography}
\end{document}